\documentclass[twocolumn,showpacs,pre,amsmath,amssymb,floatfix]{revtex4}
\usepackage{graphicx}
\usepackage{bm}
\begin{document}

\title{Volatility, Persistence, and Survival in Financial Markets}
\author{ M. Constantin}
\address{
Condensed Matter Theory Center,
Department of Physics, University of Maryland, College Park, 
Maryland 20742-4111
}
\author{ S. \surname{Das Sarma}}
\address{
Condensed Matter Theory Center,
Department of Physics, University of Maryland, College Park, 
Maryland 20742-4111
}

\begin{abstract}
We study the temporal fluctuations in time--dependent stock prices
(both individual and composite) as a stochastic phenomenon using
general techniques and methods of nonequilibrium statistical
mechanics. In particular, we analyze stock price fluctuations as a
non--Markovian stochastic process using the first--passage statistical
concepts of persistence and survival. 
We report the results of empirical measurements of the normalized 
$q$-order correlation functions $f_q(t)$, survival probability $S(t)$, 
and persistence probability $P(t)$ for several stock market dynamical 
sets. We analyze both minute--to--minute and higher frequency stock 
market recordings (i.e., with the sampling time $\delta t$ of the 
order of days). We find that the fluctuating stock price is
multifractal and the choice of $\delta t$ has no effect 
on the qualitative multifractal behavior displayed by the 
$1/q$--dependence of the generalized Hurst exponent $H_q$
associated with the power--law evolution of the correlation function 
$f_q(t)\sim t^{H_q}$. The probability $S(t)$ of the stock price
remaining above the average up to time $t$ is very
sensitive to the total measurement time $t_m$ and the sampling time. 
The probability $P(t)$ of the stock not returning to 
the initial value within an interval $t$ has a universal power--law 
behavior, $P(t)\sim t^{-\theta}$, with a persistence exponent $\theta$ 
close to $0.5$ that agrees with the prediction $\theta=1-H_2$. The 
empirical financial stocks also present an interesting feature 
found in turbulent fluids, the extended self--similarity.
\end{abstract}


\maketitle

\section{Introduction}
\label{intro_stock}
%
The financial stocks are complex, nonlinear, open systems characterized 
by a large number of parameters. Among other features, they 
present a {\it multifractal} behavior
\cite{mf1,mf2,mf3,mf4,mf5,mf-korean,mf-korean2,MFM1,MFM2,Bacry:PRE,Bacry:2001}. 
Whether or not the multifractality is intrinsic or apparent is still an 
open question \cite{JPB}. 
Several interesting multifractal models \cite{MFM1,MFM2} have been 
developed over the last decade. For example, the multifractal random 
walk model, introduced by Barcy and collaborators \cite{Bacry:PRE,Bacry:2001} 
has been recently shown to explain, besides the multifractality, other 
features of financial time--series, such as the absence of correlations 
between price variations and long-range volatility correlations. 
Other multifractal models have been proposed \cite{Pochart:preprint,Eisler:2004} 
to account for the asymmetry of the volatility-return correlation function.
An important step toward a better understanding of such intricate 
dynamical processes is to 
search for new methods that are able to provide information about 
their temporal evolution. One way to explore the temporal evolution 
of a stochastic system such as a fluctuating stock price, denoted 
by $x(t)$, is by measuring the {\it persistence probability, $P(t)$}. 
That is the probability of the stochastic variable $x(t)$ not 
reaching its original value corresponding to the starting time 
$t_0$ up to a later time $t_0+t$. This concept, closely related to the
first--passage probability, has been successfully 
implemented in surface growth phenomena \cite{krug,magda1} and has been 
used to determine the universality class and the nonlinear features 
of the underlying dynamical process through the exponent $\theta$
associated with the power--law decay $P(t)\sim t^{-\theta}$ of the
persistence probability at large times. Alternatively, one is
interested in measuring the survival probability $S(t)$ which is
defined as the probability of the stock price remaining above a
reference value up to time $t$. In contrast with the persistence
probability, we show that the survival probability depends
independently on both the total measurement time $t_m$ and the time
between successive recordings, i.e., the sampling time $\delta t$. 
The concepts of persistence and first--passage time have been 
recently introduced in econophysics \cite{Zheng,Simonsen}. 
Despite the capacity of first--passage
statistical tools to predict the degree of performance of a given
stock (i.e., how long a stock remains above a certain value, what is
the first time when it reaches a particular level), it is quite
intriguing that their use in understanding the evolution of financial
markets is rather scarce. Our study emphasizes the potential of the
survival and persistence probabilities in any time series 
investigations and shows how these concepts can be connected 
to the traditional analysis \cite{barabasi} based on the time 
evolution of price--price correlation functions. We also point 
out the {\it extended self--similar} behavior manifested by 
the stock price correlation functions. Extended self--similarity 
was originally observed in fluid turbulence problems 
\cite{turb1,turb2} and subsequently in discrete stochastic surface 
growth models \cite{Patcha}. Therefore we combine features from very 
different fields, such as surface growth, econophysics of stock 
markets, and fluid dynamics in an effort towards understanding 
the temporal evolution of financial stocks.

Our statistical study of stock market temporal fluctuations is
motivated primarily by the availability of the huge amount of
quantitative data on the stock market prices both for individual
stocks as well as for aggregate stock indices and both for
instantaneous minute--to--minute price fluctuations as well as
large--scale fluctuations over several years. Such detailed and
precise quantitative information about the time dependence of a
far--from--equilibrium stochastic process is not easy to find in real
physical systems. For example, temporal thermal fluctuations of steps on
solid surfaces, which we have recently analyzed 
\cite{magda_Pts,dan_PRE,survival} to study the persistence and
survival properties of equilibrium step fluctuations, usually provide
reliable time dependent data only over a couple of decades. Similarly,
experimental studies \cite{krugrev} of kinetic surface roughening in
interface growth, which have served as model problems for developing
the concepts of dynamical scaling in nonequilibrium growth phenomena,
usually have reliable data for a couple of decades in growth time. The
availability of extensive quantitative information on stock prices,
including minute--to--minute price fluctuations within a single day as
well as prices spanning over years (or even decades for some stocks and
index averages), therefore provides an almost unique opportunity to
statistically analyze a time--dependent non--Markovian stochastic
phenomenon using actual ``experimental'' data covering many decades in
time, in contract to the corresponding experimental studies of real
physical systems which span only a few decades in time. It should
consequently be possible to obtain definitive information about the
persistence and survival behavior of stock price fluctuations studied
as a non--Markovian stochastic phenomenon. The availability of extensive
stock price data should also allow for a detailed multifractal
analysis of stock price fluctuations, which could serve, in principle,
as a quantitative measure of ``volatility'' in the stock market since
multifractality is directly related to intermittency and intermittency
arises from the finite probability for very large scale fluctuations. 

Our work presented in this article should therefore be thought of as a
detailed empirical ``experimental'' analysis of a non--Markovian
dynamical stochastic process, namely the stock market, from the
specific perspectives of first--passage statistics (i.e., persistence
and survival probabilities and exponents) and intermittency
(i.e., multifractality and volatility). Our motivation for studying
this specific problem is the existence of huge amount of experimental
data (i.e., stock prices) of very wide dynamical range. 
The fact that there is widespread current
interest in the statistical physics community in the subject of
``econophysics'', which is nothing but the study of economics 
using the principles and methodologies of statistical mechanics, 
gives our current work some
broad context, but our own interest is, however, complementary -- we
are using the vast amount of available dynamical data on stock price
fluctuations to carry out an ``experimental'' study of the important
first--passage statistical concepts of persistence and survival in
non--Markovian stochastic phenomena. The current work is, in some
sense, a continuation of our earlier work on understanding various
stochastic phenomena (i.e. thermal step fluctuations
\cite{magda_Pts,dan_PRE,survival} and nonequilibrium surface growth
\cite{magda1}) from the first--passage statistics perspective -- here
the stochastic process under consideration being an economic
phenomenon (i.e. stock price fluctuations) rather than physical
phenomena as in the past.

Our study of the multifractal character of price fluctuations
is based on a multifractal version \cite {MF-DFA} of the traditional 
detrended fluctuation analysis \cite{DFA}. We also use a 
standard dynamical scaling analysis, inspired from surface growth 
phenomena, to show the multifractal behavior of the financial stocks, 
quantitatively extracted from the $q$--order price--price correlation
functions. Our results go in parallel with earlier analyses of 
other groups \cite{mf1,mf2,mf3,mf4,mf5,mf-korean,mf-korean2} 
which succeeded in showing multifractality in several stock markets 
and commodities. On the other
hand we investigate the persistence of price fluctuations described 
by the persistence exponent $\theta$. 
The bridge between these two analyses is provided
by the second order Hurst exponent $H_2$ associated with the correlation
function of the stock price, which has been shown \cite{krug,magda1} 
to be simply related to the persistence exponent through $H_2=1-\theta$. 
In this study we verify that this simple relation is satisfied
by both low and high frequency fluctuating financial stocks.

The data we use for our stochastic study comprise 
the daily Intel (INTC) stock
value between January 1990 and December 2002 and the composite NYSE index 
(i.e., NYA index) between January 1966 and December 2002. This
corresponds to 3355 and 9312 data points, respectively. We also
analyze sets of data recorded every minute for the Johnson and Johnson
stock (JJ 2000) and every five minutes for the Intel stock (INTC 2000) 
during the year of 2000 \cite{chr1}. This corresponds to 98280 and
19656 data points, respectively. The daily recorded stocks have an 
obvious exponential increase of their prices over several years. 
Therefore the exponential drift of 
the background can be subtracted from the stock price and we define 
a new stochastic variable $\tilde x(t)=x(t)-x_b(t)$, 
where the stochastic variable $x_b(t)$ associated with the evolution
of the background depends on  two parameters $a$ and $b$, 
$x_b(t)=a\exp(bt)$ \cite{bckgr}. We analyze the persistence 
probability of both $x(t)$ and $\tilde x(t)$. Changing the background
subtraction within reasonable limits does not affect our statistical
conclusions about persistence exponents and/or multifractality.

The rest of the paper is organized as follows. 
In Sec.~\ref{bckgr}, we define the
various dynamical correlation functions and related statistical
quantities, as well as the various exponents to be used throughout our
statistical analyses -- Sec.~\ref{bckgr} is important in introducing the
methodology of our analyses; in Sec.~\ref{RD}, we present our
extensive results and discussions; our concluding remarks are exposed
in Sec.~\ref{concl}. 

\section{Nonequilibrium statistical mechanics techniques}
\label{bckgr}
%
\subsection{Price--price correlation functions and extended 
self-similarity}
\label{FQT}
The generalized $q$--order price--price correlation function is 
defined as 
\begin{equation}
G_q(t)=\langle |x(t_0+t)-x(t_0)|^{q}\rangle^{1/q},
\label{Fqt}
\end{equation}
\noindent where $x(t)$ is the stock price and the average is over 
all the initial times ${t_0}$. $G_q(t)$ has a power--law behavior
\begin{equation}
G_q(t) \sim t^{H_q},
\label{Hq}
\end{equation}
\noindent which defines the exponent hierarchy $H_q$, also called 
the generalized Hurst exponent. The price evolution is $multifractal$ 
if the exponent hierarchy $H_q$ varies with $q$, otherwise is
fractal (in the theory of surface dynamical scaling referred to as
multi-affine and self-affine, respectively). In particular, for $q=2$,
we recover the fractional Brownian motion case described by the
well--known Hurst exponent, $0<H_2<1$. A simple way of assigning the
presence of multifractality in a stochastic stock market is by looking
at the multifractal spectra, $\tau_q=q H_q-1$. For fractals $\tau_q$
depend linearly on $q$. A nonlinear behavior of $\tau_q$ vs~$q$ is 
considered a manifestation of multifractality.

The {\it temporal} behavior in Eq.~(\ref{Hq}) of the generalized 
price--price correlation functions is analogous to the {\it spatial} 
behavior of the $q$--order structure functions of turbulent fluids or 
the generalized height difference correlation functions, 
$C_q({\bf r},t)=\langle |h({\bf x}+{\bf r},t)-h({\bf x},t)
|^q\rangle^{1/q} \sim r^{\xi_q}$ (for small distances $r$),
corresponding to surface growth models. The multiscaling behavior 
is revealed by the $q$--dependence of the scaling exponents $\xi_q$. 
The extended self--similarity (ESS) behavior in both turbulent 
fluids and surface roughening models refers to the enhancement 
of the scaling region once the $\mbox{log}|C_m(r)|$ is plotted 
against $\mbox{log}|C_n(r)|$, where $m$ and $n$ are two different 
positive integers \cite{Patcha}. In this study we show 
that for different financial stocks $G_m(t)$ depends 
on $G_n(t)$ in a power--law fashion and, although the temporal 
scaling domain is not necessarily enhanced as in the case of 
turbulent fluids or roughening models, the ESS behavior is clearly
identified and this interesting feature of the financial stocks can 
be further exploited to understand the associated multifractal 
character.

\subsection{Persistence Exponent of Fractional Brownian
  Motion}
\label{pers_fBm}
The fractional Brownian motion (fBm) is one of the simplest stochastic
models that can be used to model financial stocks or any time series
with long--range memory. Before proceeding further, we summarize the 
definition and the known first--passage property of a fBm. We then 
describe the multifractal detrended fluctuation analysis (MF-DFA) 
method which is used to estimate the generalized Hurst exponent. 

A stochastic process $\xi(t)$ 
(with zero mean $\langle \xi(t)\rangle=0$) is called a fBm if its 
two--time correlation function 
$C(t_1,t_2)= \langle[\xi(t_1)-\xi(t_2)]^2\rangle$ is 
(i) {\em stationary}, i.e., depends only on the time difference 
$|t_2-t_1|$ and (ii) grows asymptotically as a power law \cite{MV}
\begin{equation}
C(t_1,t_2)\sim |t_2-t_1|^{2H}, \quad\quad |t_2-t_1|>>1.
\label{fbm1}
\end{equation}
The parameter $0<H<1$ is called the Hurst exponent that characterizes 
the fBm and $\langle \cdots \rangle$ denotes the expectation 
value over all realizations of the process $\xi(t)$. In order to match
the notation throughout the paper, let us call the Hurst exponent
$H_2$ instead of $H$. For the sake 
of completeness we also mention that, alternatively, a zero mean 
stochastic process $\xi(t)$ is called a fBm if its autocorrelation 
function has the following expression:
\begin{equation}
a_{\xi}(t_1,t_2)=\langle \xi(t_1)\xi(t_2) \rangle 
\sim t_1^{2H_2}+t_2^{2H_2}-|t_2-t_1|^{2H_2}.  
\label{fbm2}
\end{equation}
The zero crossing 
properties of a fBm have been studied extensively in the past 
\cite{berman,hansen,ding}. In particular, assuming that $\xi(t=0)=0$, 
we are interested in the probability $P(t)$ that a fBm does $not$ 
cross zero up to time $t$ (i.e., the persistence probability): 
\begin{equation}
P(t)=\mbox{Prob} \left\{\xi(t')>0, ~\forall~ t'<t \right\}.
\end{equation}
In terms of the stochastic stock price variable $x(t)$, characterized
by a particular value $x(t_0)$ at the initial time $t_0$, the
probability of remaining always $above$ that value up to
time $t_0+t$ (i.e., $positive$ persistence) reads
\begin{equation}
P_+(t)=\mbox{Prob} \left\{x(t_0+t')>x(t_0), ~\forall~ t'<t \right\},
\end{equation}
and, similarly, the $negative$ persistence probability reads 
\begin{equation}
P_-(t)=\mbox{Prob} \left\{x(t_0+t')<x(t_0), ~\forall~ t'<t \right\}.
\end{equation}

These definitions can alternatively be reformulated in terms of 
the $cumulative$ time series of the discretized log--returns. 
Let $r_j=\mbox{ln}\left[x_{j+\Delta j}/x_j\right]$ be the discrete set
of log--returns, with $j=0,1,\ldots$. 
The sampling time $\Delta j$ is the interval between 
successive measurements. We define the cumulative log--returns, 
$R_j=\sum_{i=0}^{j}r_i$. Since $R_j=\mbox{ln}\left[x_j/x_0\right]$,
the definition of the positive persistence probability, for example,
becomes:
\begin{equation}
P_+(N)=\mbox{Prob} \left\{R_n>0, \forall~ 0 \le n <N \right\}.
\end{equation}
 
In several studies of $linear$ surface growth models, characterized by
identical positive and negative persistence probabilities, it has been 
shown that $P(t)$ decays as a power--law \cite{krug,magda1} 
at large times, $P(t)\sim t^{-\theta}$, with the steady state 
persistence exponent $\theta$ obeying the relation 
\begin{equation}
\theta=1-H_2.
\label{thetas}
\end{equation}
We note that this relation holds for any zero mean process (not
necessarily Gaussian \cite{hansen,maslov}) that satisfies the 
requirements (i) and (ii) above. Both analytic arguments as well 
as numerical simulations supporting the relation (\ref{thetas}) 
have been presented previously in the literature in the context of
fluctuating interfaces. In this study we investigate the behavior of
$P(t)$ at large times and its dependence on the
sampling time for both $x(t)$ and $\tilde x(t)$ stochastic variables. 

The persistence probability can be generalized \cite{dornic,magda_Pts}
using the persistent large deviations probability, $P(t,s)$, defined 
as the probability for the ``average sign'' $S_{\text{av}}$ of the 
stock price fluctuation to remain above a certain pre-assigned value 
``$s$'' up to time $t$:
\begin{equation}
\label{eq4}
P(t,s) \equiv \hbox{Prob}~\lbrace ~S_{\text{av}}(t^{\prime}) \geq s,
~\forall t^{\prime} \leq t~ \rbrace,
\end{equation}
\noindent where 
\begin{equation}
\label{eq5}
S_{\mbox{av}}(t) \equiv t^{-1} \int_{0}^{t} {dt^{\prime}~\hbox{sign}
~[x(t_{0} + t^{\prime}) - x(t_{0})]}.
\end{equation}
Since $S_{\text{av}}(t) \in [-1,1]$, the probability $P(t,s)$ is 
defined for $-1 \le s \le 1$. For $s=1$ we recover our earlier simple 
definition of persistence, while for $s=-1$ the probability $P(t,s=-1)$ 
is trivially equal to unity for all $t$. However, for the remaining
values of the average sign parameter $s$, $-1<s<1$, the generalization of the
persistence probability provides new information through the family of
persistent large deviations exponents, $\theta_l$, associated with 
the power--law behavior, $P(t,s)\sim t^{-\theta_l(s)}$, at large time 
scales. 

\subsection{Survival Probability}
\label{surv}
Perhaps of more practical interest in evaluating the temporal trend of
a financial stock is the probability of the stock's
price remaining above (below) a certain reference value up to a 
later time $t_0+t$, given that its initial value at time $t_0$ was
above (below) that reference level, i.e., the positive (negative) 
survival probability $S_{\pm}(t)$. Let us denote by $S(t)$ the average
between the positive and negative survival probabilities. This 
statistical quantity offers a better picture of the likelihood of 
a given stock having a positive evolving trend, for example, 
with respect to a preassigned reference value. The definition 
of this probability reads:
\begin{equation}
S_+(t)=\mbox{Prob} \left\{x(t_0+t')>\bar x, ~\forall~ t'<t \right\},
\end{equation}
where $\bar x$ is the reference price. For simplicity we consider
$\bar x$ to be the average price over the measurement time, $t_m$, but
in general it can take any value between the minimum and the maximum
values of $x(t)$ for all the discrete times $t$ up to the final
measurement time. We will show that $S(t)$ depends independently on
both $t_m$ and $\delta t$ and the scaling with $t/\delta t$ appears
only when $\delta t/t_m$ is a constant. The same type of behavior has
been found recently in experimental thermal fluctuations of 
surface steps on Ag(111), screw dislocations on the facets of Pb 
crystallites and Al--terminated Si(111) surfaces \cite{survival,dan_PRE}.

\subsection{Multifractal Detrended Fluctuations Analysis (MF-DFA)}
\label{dfa}
MF-DFA is a reliable method for analyzing correlated time series. 
It is known to provide the accurate values of the generalized Hurst
exponents even for time series with small length, while other
similar methods, such as the Hurst Rescaled Range Analysis \cite{RRA},
overestimate those values in the case of small size series 
\cite{Costa}. 

Let $x_i$, $i=1,\ldots,T_f$ be the stochastic price variable
recorded at discrete times $i$. The final transaction time is 
denoted by $T_f$. We denote by $r_j$ the log--return price variable,
$r_j=\mbox{ln}(x_{j+1}/x_j)$, $j=1,\ldots,T$, where $T=T_f-1$. We 
estimate the cumulative time series of the log-return price variables, 
\begin{equation}
X(i)=\sum_{j=1}^{i}(r_j-\overline{r}),~~~~ i=1,\ldots,T,
\end{equation}
\noindent where $\overline{r}=1/T\sum_{i=1}^{T}r_i$ is the
average value of the log-returns. The time series $X(i)$ is 
divided into $N_\tau$ disjoint segments $I_n$ ($n=1,\ldots,N_\tau$) 
of equal size $\tau$. Obviously, $N_\tau=[T/\tau]$. For each segment 
we calculate the local trend using a linear least--squared fit
$Y_{\tau}(n,t)=a_n+b_n~t$, where $t \in I_n$ and
$n=1,\ldots,N_\tau$. The local time series of the cumulative
log--returns is simply $X_{\tau}(n,t)=X[(n-1)\tau+t]$. Therefore, 
the variance is given by
\begin{equation}
F^2(n,\tau)=\frac{1}{\tau}\sum_{t=1}^{\tau}(X_{\tau}(n,t)-
Y_{\tau}(n,t))^2.
\label{F2}
\end{equation}
$F(n,\tau)$ is called the fluctuation function. In order to avoid
disregarding some data points $X(i)$ when the length $T$ of the time
series is not a multiple of the time lag $\tau$, one has to repeat
these steps starting from the opposite end of the interval. In that
case, $X_{\tau}(n,t)$ in Eq.~(\ref{F2}) becomes equal to 
$X[T-(n-N_\tau)\tau+t]$, for $n=N_\tau+1,\ldots,2N_\tau$. By 
averaging over all the segments $I_n$ we finally obtain the
correlation function of order $q$,
\begin{equation}
F_q(\tau)=\left\{\frac{1}{2N_\tau} \sum_{n=1}^{2N_\tau}
[F^2(n,\tau)]^{q/2}\right\}^{1/q}.
\end{equation}

By construction, since we use a 
linear fit for simplicity, $F_q(\tau)$ is
defined for $\tau \ge 3$. The scaling form of the correlation 
function $F_q(\tau)\sim \tau^{H_q}$ provides the family of 
generalized Hurst exponents, $H_q$. For reasons that will 
become clearer very shortly we also introduce the dimensionless 
fluctuation function, $f(n,\tau)$, defined by 
\begin{equation}
f(n,\tau)=\frac{[F^2(n,\tau)]^{1/2}}{\sigma}, 
\end{equation}
\noindent where $\sigma=\sqrt{1/T \sum_{t=1}^T (r_t-\overline{r})^2}$ 
is the standard deviation of the log-returns during the interval
$T$. Therefore, the dimensionless $q$th order correlation function
becomes
\begin{equation}
f_q(\tau)=\left\{\frac{1}{2N_\tau}\sum_{n=1}^{2N_\tau}
[f(n,\tau)]^{q}\right\}^{1/q}.
\label{fq}
\end{equation}
Obviously, $f_q$ obeys the same scaling relation as $F_q$, 
\begin{equation}
f_q(\tau)=C_{H_q} \tau^{H_q},
\label{fq_scaling}
\end{equation}
where $C_{H_q}$ is a constant independent of the time lag
$\tau$. However, for $q=2$ (which corresponds to the usual DFA
procedure), the expression of this constant is known
exactly \cite{taqqu},
\begin{equation}
C_{H_2}=\left( \frac{2}{2 H_2+1}+\frac{1}{H_2+2}-\frac{2}{H_2+1}
\right)^{1/2},
\label{CH2}
\end{equation}
where $H_2$ is the Hurst exponent of the fBm. The time evolution of
$f_q(\tau)$, along with the analytical result for the coefficient
$C_{H_2}$ can be used to understand the dynamics and memory of financial
stocks. 

\section{Results and Discussions}
\label{RD}
\begin{figure}
\includegraphics[height=12cm,width=8cm]{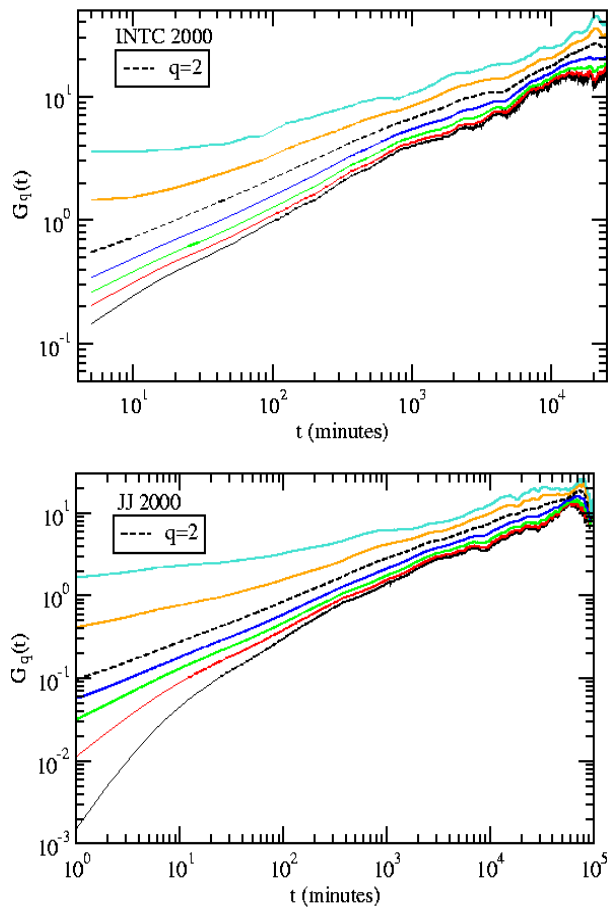} 
\caption{\label{fig1}(Color online) 
Log--log plot of the generalized price--price 
correlation function $G_q(t)$ vs.~$t$ corresponding to
minute--to--minute INTC stock (top) and JJ stock (bottom) for 
$q=1/8,1/4,1/2,1,2,4,8$ from bottom to top in each panel.}
\end{figure}
\begin{figure}
\includegraphics[height=12cm,width=8cm]{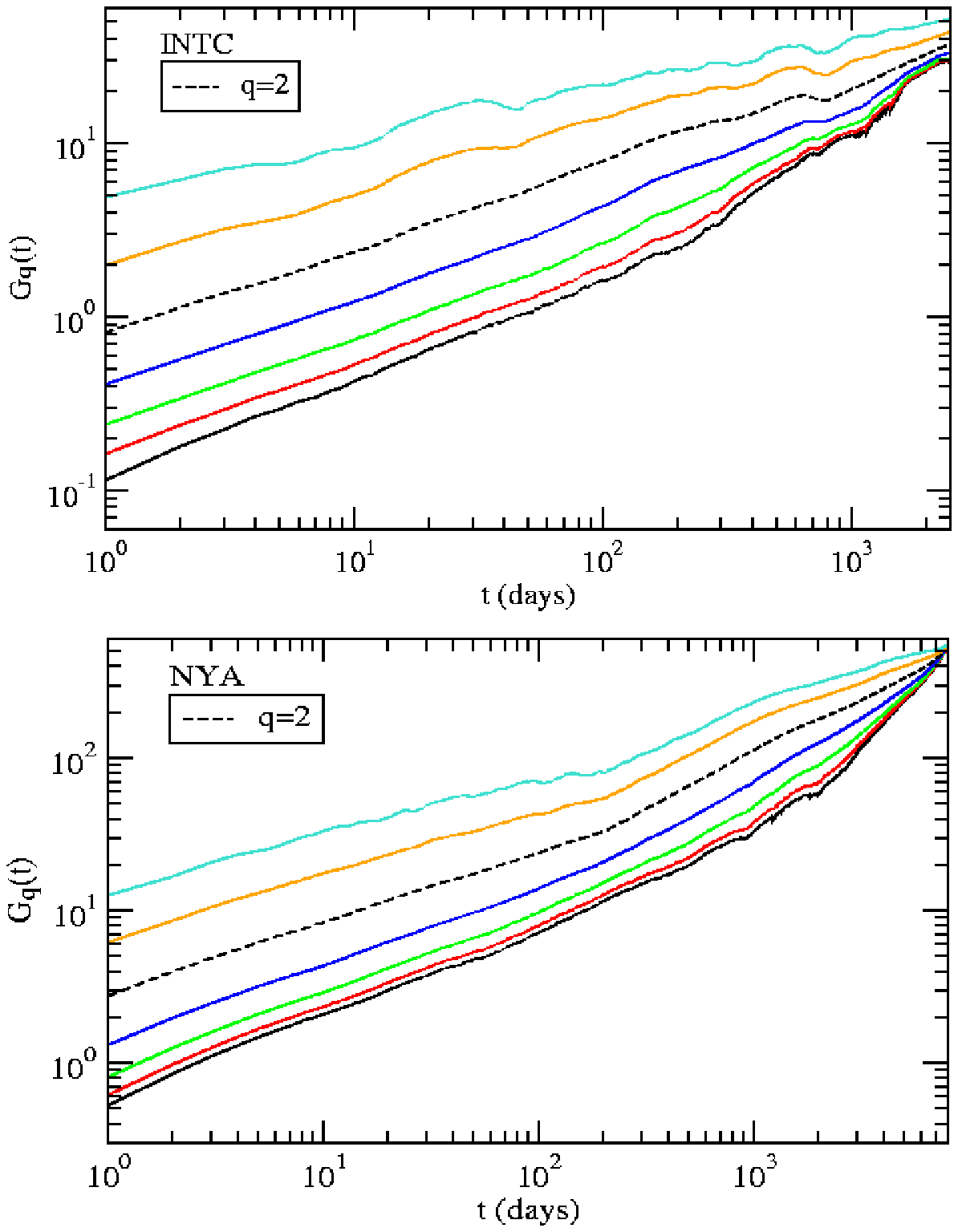}
\caption{\label{fig2}(Color online) 
Log--log plot of the generalized price--price 
correlation function $G_q(t)$ vs.~$t$ corresponding to the daily 
INTC stock (top) and NYA index (bottom) for $q=1/8,1/4,1/2,1,2,4,8$ 
from bottom to top in each panel.}
\end{figure}
\begin{figure}
\includegraphics[height=6cm,width=8cm]{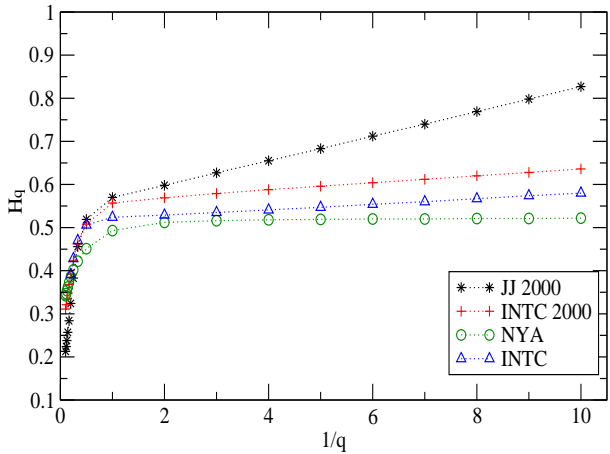}
\caption{\label{fig3}(Color online) 
The generalized Hurst exponent $H_q$ vs. $1/q$ 
for the four stocks discussed in the paper. $H_q$ behaves linearly 
with $1/q$ at both small and large values of $1/q$. For the daily
stocks a linear least--square fit has been applied to the first two
decades, while for the high frequency stocks, the fitted regions were
$25<t<500$ for the INTC 2000 stock and $25<t<400$ for the JJ 2000 
stock.}
\end{figure}

We first discuss the results concerning the price--price
correlation functions calculated using Eq.~(\ref{Fqt}). 
In Fig.~\ref{fig1} we present the results of
$G_q(t)$ for the high frequency stocks and in Fig.~\ref{fig2} for the
low frequency stocks. It is obvious that since the
log--log plots of $G_q(t)$ vs. $t$ do not exhibit linear behavior over
the entire time range the associated Hurst exponent $H_q$ varies with
time and the scaling of the correlation functions suffers many
transient regimes. A good power--law dependence appears for
$q=2$. Although for other values of $q$ the deviations from power--law
become visible it is clear that $H_q$ decreases with $q$. For
illustration purpose we have fitted certain portions of these log--log
plots to obtain a qualitative view of the dependence of $H_q$ 
on $1/q$, as shown in Fig.~\ref{fig3}. In this figure we have used a
large range of values for $q$ (i.e., $q=1/10,1/9,...,1,2,...,10$). 
We notice that $H_q$ depends
linearly on $1/q$ for both small and large values of $1/q$. The stock
with the smallest sampling time of $\delta t=1$ minute (JJ 2000) 
displays an increase of $H_q$ at large values of $1/q$, 
while for the rest of the stocks $H_q$ has the tendency to saturate 
as $1/q$ increases.  

\begin{figure}
\includegraphics[height=12cm,width=8cm]{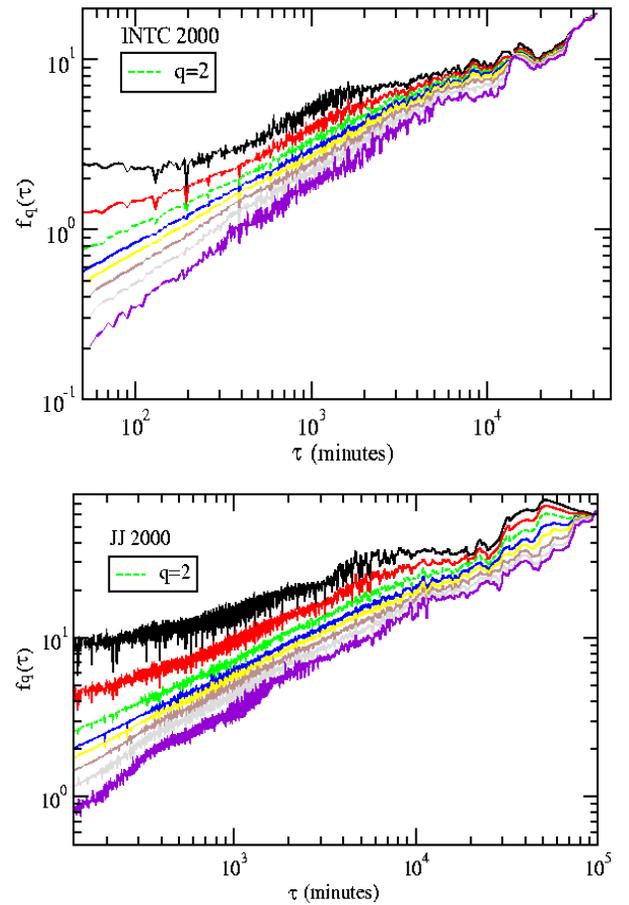}
\caption{\label{fig4}(Color online) 
Log--log plot of the normalized fluctuation 
function $f_q(\tau)$ vs.~time lag $\tau$ corresponding to INTC 2000
stock (top) and JJ 2000 stock (bottom). The MF-DFA method has been 
used. The curves shown correspond to $q=-8,-4,-2,-1/2,1/2,2,4,8$ 
from bottom to top in each panel.}
\end{figure}

\begin{figure}
\includegraphics[height=12cm,width=8cm]{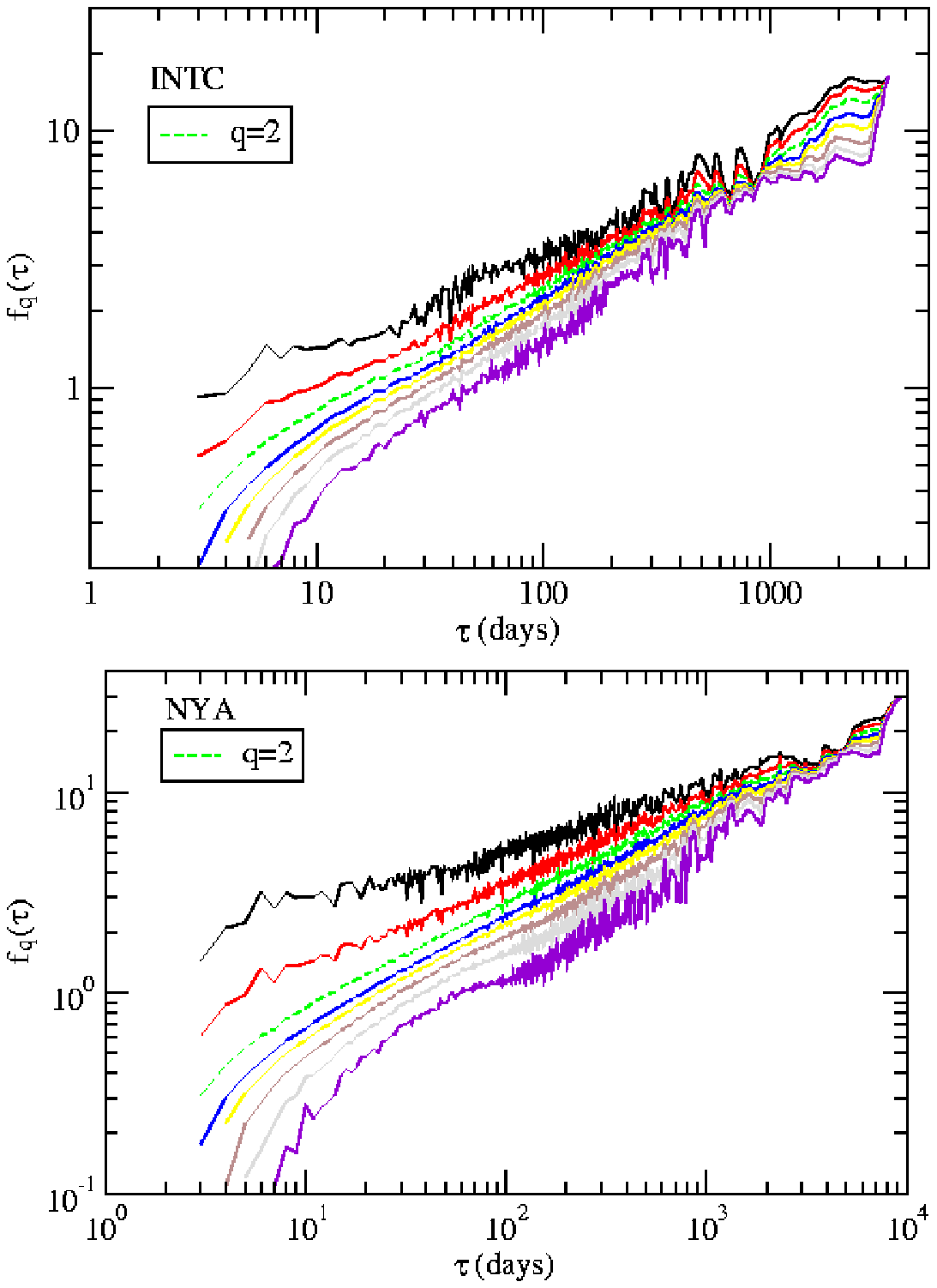}
\caption{\label{fig5}(Color online) 
Log--log plot of the normalized fluctuation 
function $f_q(\tau)$ vs.~$\tau$ corresponding to the daily INTC 
stock (top) and NYA index (bottom). The MF-DFA method has been 
used to calculate $f_q(\tau)$. The curves shown correspond to 
$-8,-4,-2,-1/2,1/2,2,4,8$ from bottom to top in each panel.}
\vspace{1.2cm}
\end{figure}

When using the MF-DFA method to calculate the correlation functions 
of order $q$ we note that the power--law behavior of $f_q(\tau)$ 
vs. $\tau$ extends over longer time periods making the extraction 
of the exponent $H_q$ more reliable. The results are shown in 
Figs.~\ref{fig4} and \ref{fig5}. This can be easily seen in the case
of the daily NYA index and INTC stocks. The power--law is seen for
more than three decades in Fig.~\ref{fig5}, while limited power--laws
spanning two decades only are seen in Fig.~\ref{fig2}.

\begin{figure}
\includegraphics[height=5.5cm,width=8cm]{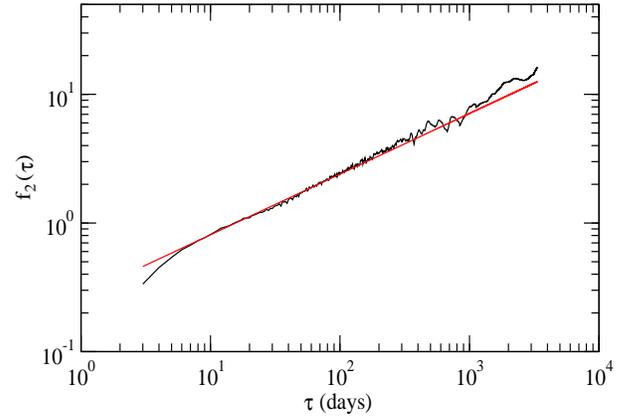}
\caption{\label{fig6}(Color online) 
The normalized fluctuation function $f_2(\tau)$ 
as a function of time for the daily INTC stock during the period 
1990--2002. The straight line represents the theoretical curve based 
on Eqs.~(\ref{fq_scaling}) and (\ref{CH2}) for $H_2=0.47$.}
\end{figure}

\begin{figure}
\includegraphics[height=5.5cm,width=7.5cm]{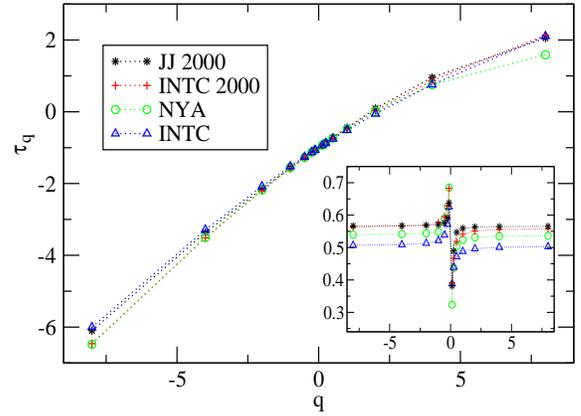}
\caption{\label{fig7}(Color online) 
The multifractal spectra $\tau_q$ vs. $q$ for
  the stocks discussed in the paper. The deviation of all curves from
  the linear dependence is a signature of the multifractal
  behavior. The inset shows the $1/q$--dependence of $H_q$, keeping
  the same symbol--stock correspondence as in the main figure.}
\end{figure}

As we have already mentioned, the value of $H_q$ is sensitive 
to the fitted region of the log--log plot of $f_q(\tau)$
vs.~$\tau$. This issue requires special treatment. Our strategy was 
to take advantage of the analytic result of Eq.~(\ref{CH2}) in order to 
identify the time window over which the second order Hurst exponent can
be extracted correctly. This can be achieved by adjusting $H_2$ 
such that the agreement between the empirical results for $f_2(\tau)$ 
using Eq.~(\ref{fq}) and the theoretical curve predicted by
Eq.~(\ref{fq_scaling}) with $q=2$ and Eq.~(\ref{CH2}) becomes 
very good. This procedure is shown in Fig.~\ref{fig6}. 
Once the time window which gives the best agreement for $q=2$ is 
identified, we use it to extract $H_q$ for a large set of $q$ 
values ($q=\pm 8,\pm 4,\pm 2,\pm 1,\pm 1/2,\pm 1/4,\pm 1/8$). The
results for the $q$--dependence of $H_q$ based on the MF-DFA method
have been used to calculate the multifractal spectra,
$\tau_q=qH_q-1$, shown in Fig.~\ref{fig7}. We observe that all
spectra deviate from a linear $q$--dependence, which is an obvious 
manifestation of the multifractality in these stocks. In the inset of
Fig.~\ref{fig7} we also show the $1/q$--dependence of $H_q$. At
positive $q$, the qualitative trend of the results is the same as in
Fig.~\ref{fig3}. It is interesting to point out that the empirical 
set with the smallest sampling time, JJ 2000, which in the case of 
the standard price--price correlation function analysis has shown 
an increasing trend of $H_q$ at large values of $1/q$, does not
present this trend anymore, $H_q$ saturating quickly as $1/q$
increases. Negative values of $q$ are accessible within the MF--DFA
analysis. For large negative values of $1/q$ the generalized Hurst
exponents saturate rather fast, as in the case of large positive
values of $1/q$.

The fact that $H_q$ (and in particular $H_2$) changes with time is a
clear indication of the multifractal character. In this context we
mention that the so--called multifractional Brownian motion could be
alternatively used to model this feature \cite{Costa}.

\begin{figure}
\includegraphics[height=5.5cm,width=8cm]{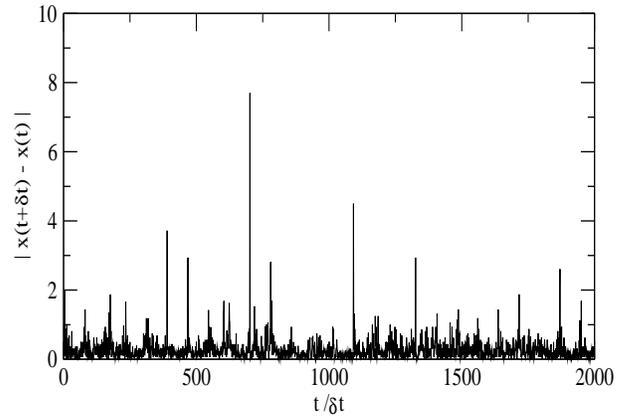}
\caption{\label{fig8} The local stock price difference 
$|x(t+\delta t)-x(t)|$ vs. the
  dimensionless variable $t/\delta t$ for the INTC 2000 recorded with
  sampling time $\delta t=5$ minutes.}
\end{figure}

\begin{figure}
\includegraphics[height=6.5cm,width=8cm]{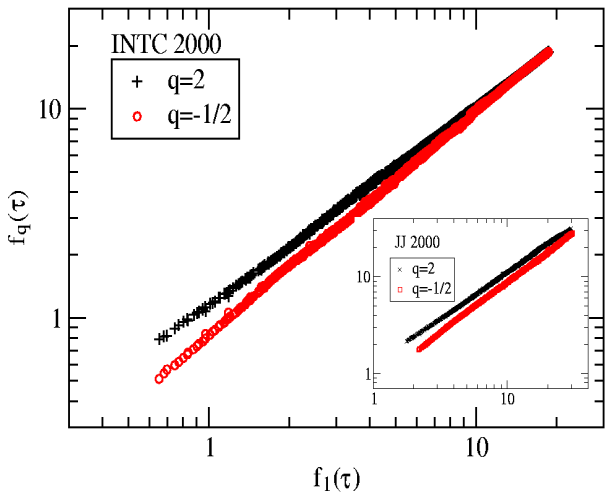}
\caption{\label{fig9}(Color online) 
The normalized fluctuation functions 
$f_2(\tau)$ and $f_{-1/2}(\tau)$ as a function of $f_1(\tau)$ 
for the INTC 2000 and JJ 2000 (inset) stocks.}
\end{figure}

The analogy of the stock market fluctuations and fluid turbulence has
already been pointed out in the literature \cite{turb_prev}. 
It is known that in turbulent fluids the energy dissipation rate shows
 violent fluctuations. Similarly, as shown in Fig.~\ref{fig8}, we find 
that the temporal evolution of the local stock price differences, 
$|x(t+\delta t)-x(t)|$, presents strong fluctuations which represent 
the signature of intermittency. 
In addition, the self-extending similarity features of intermittent 
fluid turbulence have been shown to exist in $spatial$ height
correlation functions of the kinetic surface roughening models. 
We show in Figs.~\ref{fig8} and \ref{fig9} that the extended 
self--similarity exhibited by the structure functions in fluid 
turbulence also shows up in the $temporal$ behavior of the 
financial stocks correlation functions. This observation 
offers a connection between three distinct physical problems, 
apparently without any intuitive connections: 
fluid turbulence, surface roughening and financial stocks. 

\begin{figure}
\includegraphics[height=6.5cm,width=8cm]{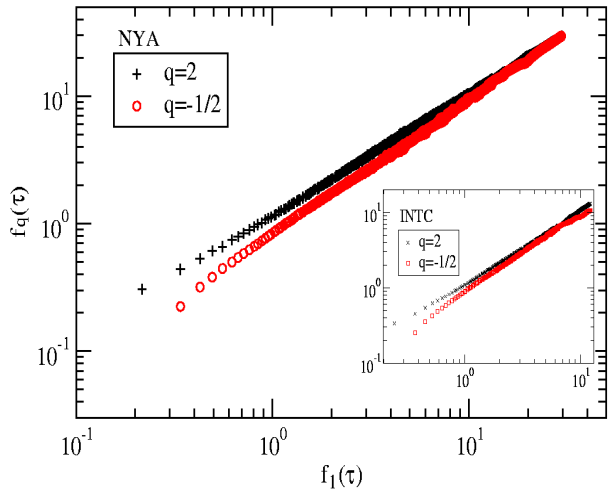}
\caption{\label{fig10}(Color online) 
The normalized fluctuation functions 
$f_2(\tau)$ and $f_{-1/2}(\tau)$ as a function of $f_1(\tau)$ 
for the daily NYA and INTC (inset) stocks.}
\end{figure}

For exemplification purpose we plot in Figs.~\ref{fig9} and 
\ref{fig10} $\mbox{log}|f_m(\tau)|$ vs. $\mbox{log}|f_n(\tau)|$ 
for $m=-1/2$ and $2$ and $n=1$. From the linear
behavior of these plots it is obvious that 
$f_m(\tau) \sim \left [f_n(\tau)\right]^{\alpha_{mn}}$, 
where the expectation value for the exponents $\alpha_{mn}$ is 
$\alpha_{mn}=H_m/H_n$. We find that our
empirical results for $\alpha_{mn}$ are in very good agreement with 
the expected ratios between $H_m$ and $H_n$, as summarized in Table 
\ref{tab1}. It would be interesting to check the existence of the
extended self--similarity in other financial stocks. We mention that
the analysis of correlation functions based on the
extended self--similarity is known to provide reliable values 
of the ratios between several generalized Hurst exponents \cite{Patcha}.  
\begin{table}
\begin{tabular}{ccccc}
\hline Stock & ~~$H_2/H_1$~~ & ~~$H_{-\frac{1}{2}}/H_1$~~ 
&~~ $\alpha_{2,1}$~~ &~~$\alpha_{-\frac{1}{2},1}$~~ \\ 
\hline \hline 
INTC & $0.966$ & $1.051$ & $0.956$ & $1.065$ \\
NYA & $0.960$ & $1.059$ & $0.961$ & $1.058$ \\ 
INTC 2000 & $0.958$ & $1.062$ & $0.956$ & $1.065$ \\ 
JJ 2000 & $0.983$ & $1.033$ & $0.962$ & $1.040$ \\ 
\hline 
\end{tabular}
\caption{Extended self-similar behavior of financial stocks based on the
  power--law dependence of $\mbox{log}|f_m(\tau)|$ on
  $\mbox{log}|f_n(\tau)|$ ($m=-\frac{1}{2}$ and $2$ and $n=1$). The
  exponents $\alpha_{-\frac{1}{2},1}$ and $\alpha_{2,1}$ extracted
  from the power--laws shown in Figs.~\ref{fig9} and \ref{fig10} are in
  very good agreement with the expected ratios of
  $H_{-\frac{1}{2}}/H_1$ and $H_2/H_1$, respectively. }  
\label{tab1}
\end{table}

\begin{figure}
\includegraphics[height=12cm,width=8cm]{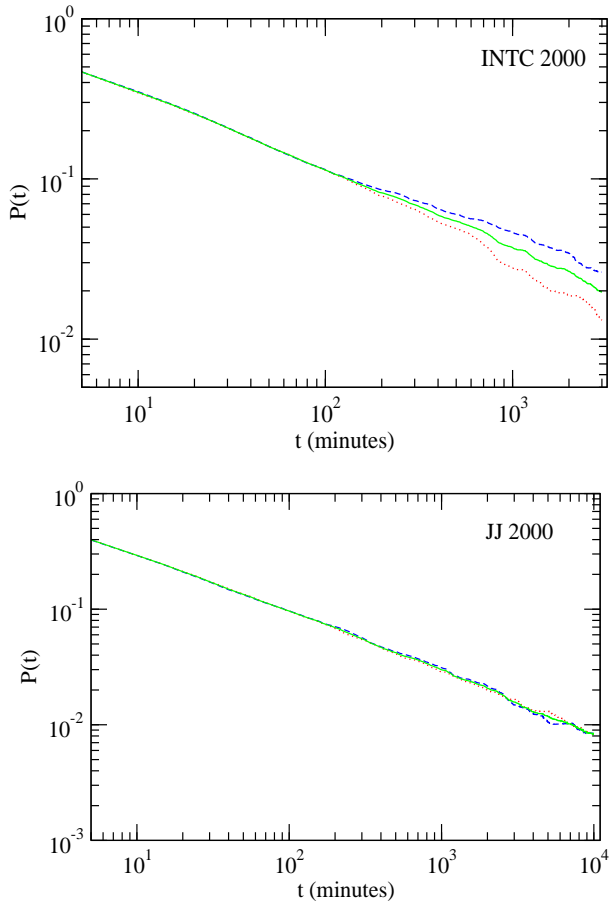}
\caption{\label{fig11}(Color online) 
Persistence probabilities $P(t)$ vs. $t$ for
  the minute--to--minute stocks. The dashed line corresponds to $P_+(t)$,
  the dotted line corresponds to $P_-(t)$, and the solid line
  represents $P(t)$. }
\end{figure}

\begin{figure}
\includegraphics[height=12cm,width=8cm]{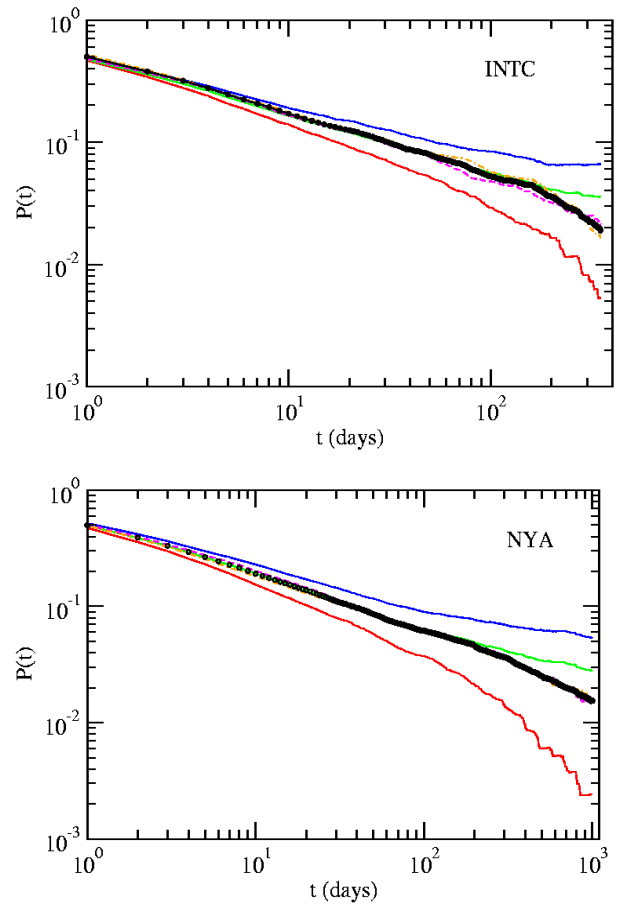}
\caption{\label{fig12}(Color online)
Persistence probabilities $P(t)$ vs. $t$ for
  the daily stocks. Solid lines correspond to $P_+(t)$, $P(t)$, and
  $P_-(t)$ (from top to bottom). The dashed lines correspond to 
$\widetilde P_+(t)$ and $\widetilde P_-(t)$. The circle represents 
the average probability $\widetilde P(t)$ for the variable 
$\tilde x(t)$. }
\end{figure}

Next, we present in the results of the persistence probabilities and
persistence exponents for our fluctuating stocks. In Fig.~\ref{fig11}
we show the results based on the minute--to--minute stocks and in
Fig.~\ref{fig12} the daily stocks. We observe that the best power--law 
appears for the average persistence probability 
$P(t)=1/2(P_+(t)+P_-(t))$, while departures from the power--law 
behavior can be seen for $P_{\pm}(t)$ corresponding to the INTC 2000 
stock and more clearly for the daily stocks. For the low frequency 
stocks, in addition to measuring the positive, negative, 
and the average persistence probabilities of the stochastic price 
variables, we have also considered the set of these three
probabilities corresponding to the empirical sets after the background 
elimination, i.e., $\tilde x(t)=x(t)-a\exp(bt)$. For the INTC
stock we have that $a=1.1703$ and $b=1.1889 \times 10^{-3}$, and for
NYA index  $a=28.473$ and $b=0.3177 \times 10^{-3}$. The persistence
curves for the variable $\tilde x(t)$ are very similar in the sense
that no distinction between the positive, negative and average
probabilities can be made. This result agrees with previous studies
of the persistence probability of the German stock index
\cite{Zheng}. We have used $P(t)$ and $\widetilde P(t)$
in order to extract the persistence exponents for the
minute--to--minute and daily stocks, respectively. The results are
summarized in Tab.~\ref{tab2}. We compared these values against
$1-H_2$, with $H_2$ extracted from the fitted power--law of
$f_2(\tau)$, in order to investigate the validity of
Eq.~(\ref{thetas}). We find a good agreement between $\theta$ and
$1-H_2$. However, it is important to emphasize that since both
$\theta$ and $H_2$ are very close to $1/2$ the memory effects of the
time series under investigation can only be revealed by higher order
correlation functions. The second order correlation function by itself
cannot explain the multifractality discussed in this study
since it indicates that the returns are uncorrelated. We also add that
$P(t)$ is not sensitive to the large discrepancy between $\delta t$ and
$t_m$ corresponding to the high frequency stocks and daily stocks,
respectively. 
\begin{table}
\begin{tabular}{ccc}
\hline Stock & $\theta$ & $H_2$ \\ 
\hline \hline INTC & $0.51$~~  & $0.47$ \\  
NYA & $0.50$~~  & $0.50$ \\ 
INTC 2000~~~ & $0.50$~~  & $0.47$ \\  
JJ 2000 & $0.52$~~ & $0.47$ \\ 
\hline
\end{tabular}
\caption{ The persistence exponent $\theta$ associated with the power--law
  decay of the average probability $P(t)$ ($\widetilde P(t)$) for the
  minute--to--minute stocks (daily stocks). $H_2$ is the second order
  Hurst exponent extracted from the time evolution of $f_2(\tau)$ (see
  Eq.~(\ref{fq_scaling})).}  
\label{tab2}
\end{table}

\begin{figure}
\includegraphics[height=6cm,width=8cm]{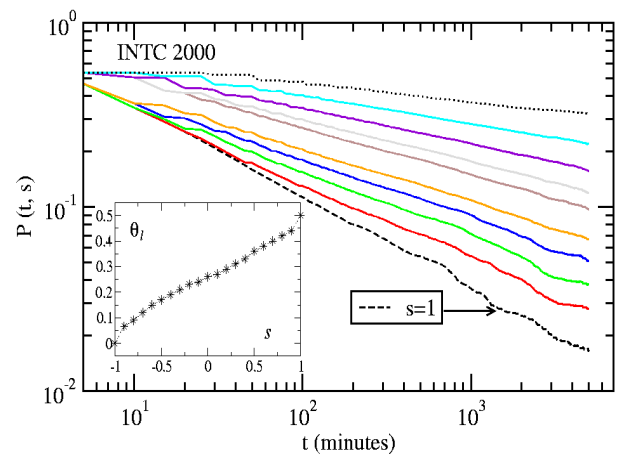}
\caption{\label{fig13}(Color online) 
Log--log plot of the persistent large
  deviations probability $P(t,s)$ vs. $t$ for
  the INTC 2000 stock. The average sign parameter, $s$, is varied from
  1 to -0.8 (bottom to top) with an increment of 0.2. The inset
  contains the persistent large deviations exponent $\theta_l$ vs. $s$.}
\end{figure}

The generalization of the persistence probability is shown in 
Fig.~\ref{fig13}. We only present the results for the probability of 
persistent large deviations corresponding to the INTC 2000 stock, but 
we have checked the applicability of this concept to other 
empirical stocks as well. From the linear behavior of
$\mbox{log}P(t,s)$ vs. $\mbox{log}t$ we conclude that the
$t$--dependence of $P(t,s)$ is indeed a power--law. We see that the 
local slope decreases as the average sign parameter $s$ decreases. 
We have varied $s$ from $-1$ to $1$ with an increment of $0.1$ and 
the $s$--dependence of the resulting family of persistent large 
deviations exponents is shown in the inset of Fig.~\ref{fig13}. We 
mention that each curve in Fig.~\ref{fig13} corresponds to the 
average between the positive and negative persistent large deviations 
probabilities, i.e., $P(t,s)=1/2(P_+(t,s)+P_-(t,s))$. Both $P_+(t,s)$ 
and $P_-(t,s)$ show departures from the expected power--laws at large 
$t$, as we have seen in Figs.~\ref{fig11} and \ref{fig12} in the 
case of the positive and negative persistence probabilities.

\begin{figure}
\includegraphics[height=12cm,width=8cm]{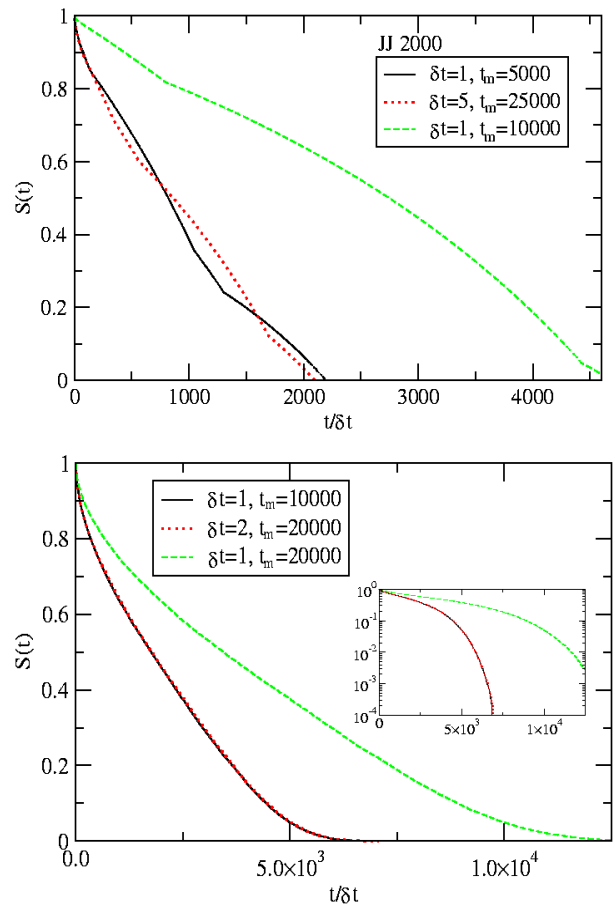}
\caption{\label{fig14}(Color online) 
The average survival probability $S(t)$ 
vs. $t/\delta t$ for the JJ 2000 stock (top) with different sampling 
times and different measurement times, as shown in the legend. 
The bottom plot contains the numerical simulation of $S(t)$ 
corresponding to a random walk with $L=100$. The scaling of 
$S(t)$ with $t/\delta t$ appears when $\delta t/t_m$ is constant. 
The inset shows the same curves on a logarithmic--linear scale.}
\end{figure}

Finally we show the results of the survival probability in
Fig.~\ref{fig14}. We have looked at the temporal evolution of the JJ
2000 stock price recorded with two different sampling times, of 1 and 5
minutes, respectively. We have chosen different values of the
measurement time, $t_m$, which influences directly the value of the
average price $\bar x=1/t_m\int_{0}^{t_m}
{dt^{\prime}~x(t^\prime)}$. As shown in upper panel of 
Fig.~\ref{fig14} the survival
probability measured over a longer $t_m$ has a slower decrease  with
time than the one corresponding to a smaller $t_m$. We find that 
the empirical measurements of $S(t)$ show good scaling with 
$t/\delta t$ at short times ($t<300$ minutes), when the ration
between the sampling time and measurement time is kept constant. This
agrees with similar measurements done on experimental step
fluctuations \cite{dan_PRE}. Since a very simple interpretation of the
stock market fluctuations is based on the random walk model, we have
numerically simulated the survival behavior of a random walker which
allows us to understand qualitatively all the features of $S(t)$ found
experimentally. Measurements of $S(t)$ for the random walk model were
carried out for systems of size $L=100$. The measured average of the
random walker variable at each site over the measurement time was used 
as the reference level in the calculation of the survival probability
and the results were averaged over 300 independent runs. 
In bottom panel of Fig.~\ref{fig14} we show that a perfect collapse
of $S(t)$ vs.~$t/\delta t$ appears when the ratio $\delta t/t_m$ is
constant. In addition, $S(t)$ corresponding to the data recorded with
the same $\delta t$ shows a slower decrease when the
measurement time is larger, as in the empirical case. Therefore both
sampling time and total measurement time have to be taken into
consideration in order to interpret correctly the survival features 
of the financial systems. We want to point out that $S(t)$ does not 
show an exponential behavior over the investigated time range 
(see the inset of the bottom panel of Fig.~\ref{fig14}), 
and possibly much larger $t_m$ is needed to observe such a behavior 
at large time scales, as it happens in the case of equilibrium 
surface step fluctuations \cite{survival}. 

\section{Conclusions}
\label{concl}
We conclude with some speculative thoughts on the possible development
of ``understanding'' {\it in the sense of physics} with respect to
stock price fluctuations. In physics, e.g. step fluctuations
\cite{krug,magda1,magda_Pts,dan_PRE,survival} or kinetic 
surface roughening \cite{barabasi}, one typically looks for minimal
(in the renormalization group sense) dynamical (in general, nonlinear)
stochastic continuum partial differential equations underlying the
stochastic phenomena, e.g., the Edwards--Wilkinson equation \cite{EW},
the Mullins--Herring equation \cite{MH}, the Kardar--Parisi--Zhang
equation \cite{KPZ}, the Villain--Lai--Das Sarma equation 
\cite{vill,LD}, etc.,
hoping to derive the long--time coarse--grained asymptotic power law
behavior of the system arising from some simple minimal underlying
dynamics (which is often based on symmetry and universality
considerations). It is unclear whether such an approach based on
continuum nonlinear stochastic equations is at all meaningful for the
understanding of the dynamical evolution of complex economic phenomena
such as stock price fluctuations. It may be possible to empirically
construct dynamical equations which are {\it sufficient} to reproduce
the exponents and the related statistical stochastic behavior
described in this paper, but the {\it necessary} conditions for
obtaining such equations are simply unknown (in fact, we do not know
if such equations exist, except in some ``trivial'' data fitting
sense). It is therefore quite intriguing that stock price fluctuation
data are amenable to stochastic analyses based on first--passage
statistics and multifractality (as carried out in this article) with
results not that dissimilar from physical processes such as step
fluctuations or nonequilibrium growth. 

Our work demonstrates that the persistence and the multifractal
behaviors of stock prices (both individual and composite) are subtle
(including extended self-similar properties not identified in the literature
before). It will be of interest to investigate if the empirical behavior we
report in this paper can be derived from the various multifractal stochastic
models \cite{Bacry:PRE,Bacry:2001,Pochart:preprint,Eisler:2004} for stock 
price fluctuations proposed in the literature. Such investigations, clearly 
beyond the scope of the current work, would, however, be quite difficult 
since both persistence and extended self--similarity are notoriously difficult 
concepts to derive theoretically, even when the underlying non--Markovian 
dynamics is known for a process. At this early stage of our understanding 
of econophysics, the fact that the stock price fluctuations seem to follow 
the persistence and the multifractal properties of well--studied surface 
fluctuation phenomena is by itself intriguing and interesting.

To summarize, in this study we have analyzed the multifractality,
extended self--similarity, and first--passage properties of
several financial stocks. While the second order Hurst exponent and the
persistence exponent characterizing the power--law decay of the
average persistence probability are not able to explain the long--term 
correlations in the investigated price time series, higher order 
correlation functions reveal much richer information about the 
complicated dynamics of such systems. We have shown that the
persistence exponent $\theta$ is in agreement with $1-H_2$ and 
does not depend on the sampling time and measurement time. However, 
the survival probability has a nontrivial dependence on both 
$\delta t$ and $t_m$, presenting scaling with $t/\delta t$ only 
when the ration $\delta t/t_m$ is a constant. The numerical 
simulations of persistence, survival, and extended 
self--similarity features using discrete models of 
financial stocks remain on interesting open problem.


The authors wish to thank A. Christian Silva and Victor M. Yakovenko 
for useful discussions and providing most of the empirical data sets 
used in this study. We are indebted to Professor Michael E. Fisher for 
his proof reading of our manuscript and his positive comments. This work 
is partially supported by the NSF and U.S. ONR.


\end{document}